\documentclass[12pt]{article}
\usepackage{amssymb,amsmath,epsfig}

\begin{document}

\title{\bf Dynamics of Charged Radiating Collapse in Modified Gauss-Bonnet Gravity}

\author{M. Sharif$^1$ \thanks{msharif.math@pu.edu.pk} and G. Abbas$^2$ \thanks{ghulamabbas@ciitsahiwal.edu.pk}
\\
$^1$Department of Mathematics, University of the Punjab,\\
Quaid-e-Azam Campus, Lahore-54590, Pakistan.\\
$^2$Department of Mathematics, COMSATS Institute\\
of Information Technology, Sahiwal-67000, Pakistan.}

\date{}
\maketitle
\begin{abstract}
This paper deals with the dynamics of a shearfree charged radiating
collapse in modified Gauss-Bonnet gravity. The field equations for
shearfree spherical interior geometry of a charged dissipative star
are formulated. To study the dynamical behavior of collapsing
matter, we derive the dynamical as well as transport equations. We
conclude that the gravitational force in modified Gauss-Bonnet
gravity is much stronger as compared to general relativity which
implies the increase in the rate of collapse. Finally, we study the
effect of charge on the dynamics of collapse.
\end{abstract}
Gravitational collapse.

\section{Introduction}

During the last few decades, there has been a growing interest to
study fate of universe in alternative theories of gravity. These
theories provides the recognition of dark energy (DE) which might be
the major agent for the rapid expansion of the universe
\cite{1a}-\cite{1d}. The consistency of such theories has been
confirmed with the observations and experiments \cite{2a}-\cite{2c}.
The most simple modification to general relativity (GR) is $f(R)$
theory of gravity in which $f$ is an arbitrary function of the Ricci
scalar $R$. Although, it is the simplest generalized form of GR but
one can formulate such general models of $f(R)$ that are consistent
with the gravitational experiments. The validity of $f(R)$ models
requires some additional restrictions due to which the theory loses
its original features \cite{3}-\cite{3b}.

The modified Gauss-Bonnet theory, so called $f(G)$ gravity has be
proposed by several authors \cite{4}-\cite{6} as an alternative
theory of gravity. The mathematical structure of this theory can be
obtained by introducing some arbitrary function $f(G)$ in
Einstein-Hilbert action of standard GR, where
$G=R^2-4R^{\mu\nu}R_{\mu\nu}+R^{\mu\nu\gamma\delta}R_{\mu\nu\gamma\delta}$
is the Gauss-Bonnet invariant. This theory is consistent with
observational constraints and might be helpful for reproducing the
cosmic history \cite{7}. It has been found that transition of the
universe form matter dominated era to accelerated phase would be
explained in the framework of $f(G)$ theory of gravity \cite{8}.
Myrzakulov et al. \cite{8a} have investigated that in this theory
$\Lambda$CDM model can be well explained without cosmological
constant term.

The dynamics of the radiating gravitational collapse is an important
issue in GR. Initially this problem was formulated by Misner and
Sharp \cite{9} for non-radiating collapse and by Misner \cite{10}
for radiating collapse. The radiating process during the formation
of neutron star or black hole \cite{11} is important due to the
energy loss. It has been investigated \cite{12,13} that
gravitational collapse is a radiating process, so the effects of
dissipation must be studied for the collapse of massive star.
Herrera and Santos \cite{14} discussed the dynamics of spherically
symmetric shearfree anisotropic dissipative fluid. Chan \cite{15}
formulated the realistic model of a dissipative star with shear
viscosity.

Herrera \cite{16} examined the role heat flux during dynamics of
radiating matter collapse. Herrera et al. \cite{17} also formulated
the dynamical equations and causal heat transport equations for
viscous and non-adiabatic collapse. They proved that the shearfree
slowly evolving and non-dissipative self-gravitating shearfree in
Newtonian limit leads to homologous collapse. Di Prisco et al.
\cite{18} investigated the dynamics of charged viscous non-adiabatic
gravitational collapse. Sharif and his collaborators
\cite{19}-\cite{25} have investigated the dynamics of charged and
uncharged dissipative gravitational collapse in GR as well as in
modified theories of gravity.

In this paper, we extend our previous work \cite{25}, to the charged
case. We use the field equations of modified Gauss-Bonnet gravity
\cite{26} derived by using the covariant gauge invariant (CGI)
perturbation approach with $3+1$ formalism. The collapsing matter
under consideration has been taken as charged radiating in the
interior of a star. The interior geometry of the star is matched
with the charged Vaidya geometry by using Darmois junction
conditions \cite{27}. This plane of the paper is as follows: In the
next section, we present the field equations in modified
Gauss-Bonnet gravity for charged radiating shearfree collapse and
discuss the matching conditions. We derive the dynamical and
transport equations and then couple them in section \textbf{3}. The
last section provides the summary of the results.

\section{Field Equations in Modified Gauss-Bonnet Gravity}

The Einstein-Hilbert action for the modified Gauss-Bonnet gravity is
\begin{equation}\label{1}
S=\int d^4x\sqrt{-g}\left(\frac{R+f(G)}{2\kappa}+\mathcal{L}_m
\right),
\end{equation}
where $g$ is the determinant of the metric tensor $g_{\mu\nu},~R$ is
the Ricci scalar, $f(G)$ is an arbitrary function of Gauss Bonnet
invariant $G,~\mathcal{L}_m$ is the matter Lagrangian and $\kappa$
is the coupling constant. The variation of this action with respect
to the metric tensor gives the field equations
\begin{eqnarray}\label{2}
G_{\mu\nu}&=&\kappa
T_{\mu\nu}+\frac{1}{2}g_{\mu\nu}f-2f_GRR_{\mu\nu}+4f_GR^{\gamma}_{\mu}R_{\nu\gamma}
-2f_GR_{\mu\gamma\delta\omega}R^{\gamma\delta\omega}_\nu\nonumber\\
&-&4f_GR_{\mu\gamma\delta\nu}R^{\gamma\delta}+2R\nabla_{\mu}\nabla_\nu
f_G-2g_{\mu\nu}\nabla^2f_G-4R^{\gamma}_{\mu}\nabla_{\nu}\nabla{\gamma}f_G\nonumber\\
&-&4R^{\gamma}_{\nu}\nabla_{\mu}\nabla{\gamma}f_G
+4R_{\mu\nu}\nabla^2f_G+4g_{\mu\nu}R^{\gamma\delta}\nabla_{\gamma}\nabla_{\delta}f_G
-4R_{\mu\gamma\nu\delta}\nabla^{\gamma}\nabla^{\delta}f_G,\nonumber\\
\end{eqnarray}
where $f_G={\partial{f(G)}}/{\partial{G}}$. The metric interior to
$\Sigma$ is assumed to be comoving and shearfree in the following
form
\begin{equation}\label{2aa}
ds^2=A^2dt^2-B^2(dr^2+r^2d\theta^2+r^2\sin^2\theta d\phi^2),
\end{equation}
where $A$ and $B$ are functions of time $t$ and radial coordinate
$r$. The corresponding four velocity, heat flux and radial vector
take the form
\begin{equation}\label{3aa}
V^\mu=A^{-1}{\delta^\mu}_{0},\quad
q^\mu=q{\delta^\mu}_{1},\quad\chi^{\mu}=B^{-1}\delta^{\mu}_{1},
\quad
\end{equation}
where $V^{\mu}q_{\mu}=0.$ The expansion scalar for the fluid sphere
is $\Theta=\frac{3\dot{B}}{AB}$.

The energy-momentum tensor for charged radiating fluid is
\begin{equation}\label{3}
T_{\mu\nu}=({\rho}+p)V_{\mu}V_{\nu}-pg_{\mu\nu}+q_\mu V_\nu +q_\nu
V_\mu +\frac{1}{4\pi}\left(-F_{\mu}^{\gamma}F_{\nu\gamma}
+\frac{1}{4}F^{\gamma\delta}F_{\gamma\delta}g_{\mu\nu}\right),
\end{equation}
where $\rho,~p,~V_\mu$,$~q_\mu$ and $F_{\mu\nu}$ are density,
pressure, four velocity, radial heat flux and anti-symmetric Maxwell
field tensor, respectively. The Maxwell equations are given by
\begin{eqnarray}\label{4ab}
F_{\mu\nu}=\Phi_{\nu,\mu}-\Phi_{\mu,\nu},\quad
{F^{\mu\nu}}_{;\nu}=4\pi J^{\mu},
\end{eqnarray}
where $\Phi_{\mu}$ is the four potential and $J_{\mu}$ is the four
current. We assume that charge is at rest with respect to comoving
coordinate system, so the magnetic field is zero. Consequently, the
four potential and the four current can be chosen as
\begin{equation}\label{6ab}
\Phi_{\alpha}=\Phi{\delta^{0}_{\alpha}},\quad J^{\alpha}=\sigma
V^{\alpha},
\end{equation}
where $\Phi=\Phi(t,r)$ is an arbitrary function and
$\sigma=\sigma(t,r)$ is the charge density. For the interior
spacetime, the Maxwell field equations take the form
\begin{eqnarray}\label{8ab}
\Phi''-\left(\frac{A'}{A}-\frac{B'}{B}-\frac{2}{r}\right){\Phi'}&=&{4
\pi }{\sigma}AB^{2},
\\\label{9ab}
{\dot{\Phi}}'-\left(\frac{\dot{A}}{A}-\frac{\dot{B}}{B}\right){\Phi'}&=&0,
\end{eqnarray}
where dot and prime represent partial derivatives with respect to
$t$ and $r$, respectively. Integration of Eq.(\ref{8ab}) implies
that
\begin{equation}\label{10}
\Phi'=\frac{sA}{Br^{2}},
\end{equation}
where $s\left(r\right)=4{\pi}{\int^{r}_{0}}\sigma B^3r^{2}dr$ is the
total charge inside the spherical symmetry.

In general, it is very difficult to deal with the field equations
(\ref{2}). We use a simplified form of the field equations derived
by Li et al. \cite{26}. For this purpose, they used the CGI
perturbation approach with $3+1$ formalism. The highly nonlinear
terms in $R_{\mu\nu}$ and $R_{\mu\nu\lambda\sigma}$ can be expressed
in the form of dynamical quantities. Consequently, the field
equations as well as other quantities are simplified as follows
\begin{equation}\label{4}
G_{\mu\nu}=\kappa(T_{\mu\nu}+T^{G}_{\mu\nu}),
\end{equation}
where $T^G_{\mu\nu}$ is the Gauss-Bonnet correction term. Using CGI
approach, the components of $T^G_{\mu\nu}$ are evaluated in terms of
the dynamical quantities as \cite{26}
\begin{eqnarray}\label{5}
{\rho}^G&=&\frac{1}{\kappa}\left(\frac{1}{2}(f-f_GG)+\frac{2}{3}(\Omega-3\Psi)
(\dot{f_G}\Theta+{\tilde{\nabla}}^2f_G)\right),\\\label{6}
{-p}^G&=&\frac{1}{\kappa}\left(\frac{1}{2}(f-f_GG)+\frac{2}{3}(\Omega-3\Psi)\ddot{f_G}-
\frac{8}{9}\Omega(\Theta
\dot{f_G}+{\tilde{\nabla}}^2f_G)\right),\\\label{7}
q^G_{\mu}&=&\frac{1}{\kappa}\left(-\frac{2}{3}(\Omega-3\Psi)(\tilde{\nabla}_\mu
\dot{f_G}-\frac{1}{3}\Theta\tilde{\nabla}_\mu{f_G})+\frac{4}{3}\dot{f_G}\Theta\zeta_\mu\right).
\end{eqnarray}
Here, $\tilde{\nabla}$ is spatial covariant derivative,
$\Theta=V^{\mu}_{~;\mu}$ is the expansion scalar and the quantities
$\Omega,~\Psi,~\zeta_\mu$ are expressed in terms of dynamical
quantities as
\begin{eqnarray}\label{8}
\Omega&=&-(\dot{\Theta}+\frac{1}{3}\Theta^2-\tilde{\nabla}^{\mu}A_\mu),\\\label{9}
\Psi&=&-\frac{1}{3}(\dot{\Theta}+\Theta^2+\tilde{R}-\tilde{\nabla}^{\mu}A_\mu),\\\label{10}
\zeta_\mu&=&-\frac{2\tilde{\nabla}_\mu \Theta}{3}+\tilde{\nabla}^\nu
\sigma_{\mu\nu}+\tilde{\nabla}^\nu \omega_{\mu\nu},
\end{eqnarray}
where $\tilde{R}$ is the Ricci scalar of $3D$ spatial spherical
surface, $\sigma_{\mu\nu}=V_{(\mu;\nu)}-A_{(\mu}V_{\nu)}-
\frac{1}{3}\Theta h_{\mu\nu}$ (where $h_{\mu\nu}=g_{\mu\nu}-V_{\mu}
V_\nu$) is the shear tensor, $A_{\mu}=V_{\mu;\nu}V^{\nu}$ is four
acceleration and
$\omega_{\mu\nu}=V_{[\mu;\nu]}+\dot{V}_{[\mu}V_{\nu]}$ is the
vorticity tensor.

The Gauss-Bonnet invariant $G$ in CGI approach is
\begin{equation}\label{11}
G=2(\frac{1}{3}R^2-{R}^{\mu\nu}R_{\mu\nu}),
\end{equation}
where
\begin{eqnarray*}\label{12}
&R&=-2\dot{\Theta}-\frac{4}{3}\Theta^2+2\tilde{\nabla}^{\mu}A_\mu-\tilde{R},\\\label{13}
&R^{\mu\nu}R_{\mu\nu}&=\frac{4}{3}(\dot{\Theta}^2+\dot{\Theta}\Theta^2
+\frac{1}{3}\Theta^4)+\frac{2}{3}(\dot{\Theta}+\Theta^2)\tilde{{R}}
-\frac{8}{3}(\dot{\Theta}+\frac{1}{2}\Theta^2)\tilde{\nabla}^{\mu}A_\mu.\nonumber
\end{eqnarray*}
Using Eqs.(\ref{3}), (\ref{4}), (\ref{2aa}) and (\ref{3aa}), the
field equations yield
\begin{eqnarray}\label{16}
8{\pi}(\rho+\rho^G)A^{2}+\frac{(sA)^2}{(rB)^4}&=&3\left(\frac{\dot{B}}{B}\right)^2
-\left(\frac{A}{B}\right)^{2}\left(2\frac{B''}{B}
-\left(\frac{B'}{B}\right)^2+\frac{4B'}{B r}\right),\\\label{17}
8{\pi}(q+q^G)AB^2&=&{2}
\left(\frac{\dot{B'}}{B}-\frac{\dot{B}B'}{B^2}-\frac{\dot{B}A'}{BA}\right),\\\label{18}
8{\pi}(p+p^G) B^{2}-\frac{(sB)^2}{(rB)^4}
&=&\left(\frac{B'}{B}\right)^{2}+\frac{2}{r}\left(\frac{B'}{B}+\frac{A'}{A}\right)+2\frac{A'B'}{AB}
-\left(\frac{B}{A}\right)^2\nonumber\\
&\times&\left(2\frac{\ddot{B}}{B}+\left(\frac{\dot{B}}{B}\right)^2
-2\frac{\dot{A}\dot{B}}{AB}\right),\\\label{19}
8{\pi}(p+p^G)r^2B^{2}+\frac{s^2}{(rB)^2}
&=&r^2\left(\frac{B''}{B}-\left(\frac{B'}{B}\right)^2+\frac{1}{r}\left(\frac{B'}
{B}+\frac{A'}{A}\right)+\frac{A''}{A}\right)\nonumber\\
&-&r^2\left(\frac{B}{A}\right)^2\left(2\frac{\ddot{B}}{B}
+\left(\frac{\dot{B}}{B}\right)^2-2\frac{\dot{A}\dot{B}}{AB}\right),
\end{eqnarray}
where $\rho^G,~p^G$ and $q^G$ correspond to Gauss-Bonnet
contribution to GR. Making use of Eqs.(\ref{5})-(\ref{7}), these
turn out to be
\begin{eqnarray}\label{20}
\kappa\rho^G&=&\frac{1}{2}(f-Gf_G)+\frac{4}{3B^2}\left(\frac{3\dot{B}^2}{A^2}-\left(\frac{B'}{B}\right)^2
+\frac{4B'}{Br}+2\frac{B''}{B}\right)\nonumber\\
&\times&\left(\frac{3\dot{f_G}\dot{B}}{AB}+\tilde{\nabla}^2f_G\right),
\\\label{21} \kappa
p^G&=&\frac{1}{2}(Gf_G-f)-\frac{4}{3B^2}\left(\frac{3\dot{B}^2}{A^2}
-\left(\frac{B'}{B}\right)^2+\frac{4B'}{Br}+2\frac{B''}{B}\right)\ddot{f}_G\nonumber\\
&-&\frac{8}{3}\left(\frac{\ddot{B}}
{AB}-\frac{\dot{A}\dot{B}}{A^2B}-\frac{\dot{B}^2}{AB^2}+\frac{\dot{B}^2}{A^2B^2}-\frac{1}{3B^2}
\left(\frac{A''}{A}-\frac{A'^2}{A^2}+2\frac{A'}{rA}+\frac{A'B'}{AB}\right)\right)\nonumber\\
&\times&\left(\frac{3\dot{f_G}\dot{B}}{AB}+\tilde{{\nabla}}^2f_G\right),
\end{eqnarray}
\begin{eqnarray}\label{22}
\kappa
q^G&=&\frac{4}{3B^2}\left(\frac{3\dot{B}^2}{A^2}-\left(\frac{B'}{B}\right)^2
+\frac{4B'}{Br}+2\frac{B''}{B}\right)\left(\frac{\dot{B}f_G'}{BA}-\dot{f_G}'\right)\nonumber\\
&+&\frac{8\dot{f_G}\dot{A}}{BA}\left(\frac{\dot{B}B'}{AB^2}+\frac{\dot{B}A'}{{A^2B}}-\frac{\dot{B'}}{AB}\right).
\end{eqnarray}

The Gauss-Bonnet invariant is found from Eq.(\ref{11}) as
\begin{eqnarray}\nonumber
G&=&\frac{2}{3}\left(-6
\left(\frac{\dot{B}}{AB}\right)^{.}-12\frac{\dot{B}^2}{(AB)^2}-\frac{2}{B^2}
\left(\frac{A'B'}{AB}-\left(\frac{A'}{A}\right)'\right)\right.\\\nonumber
&+&\left.4\frac{A'(Br)'}{AB^3
r}-2\left(\frac{4B'}{{B^3r}}+2\frac{B''}{B^3}-\frac{B'^2}{{B^4}}\right)\right)^2-{24}\left
(\left(\frac{\dot{B}}{(AB)}\right)^{.}\right)^2\nonumber\\
&-&\frac{8}{3}\left(\frac{3\dot{B}}{AB}\right)^2\left(\left(\frac{3\dot{B}}{AB}\right)^{.}+\frac{1}{3}
\left(\frac{3\dot{B}}{AB}\right)^2\right)-\frac{8}{3}
\left(\left(\frac{3\dot{B}}{AB}\right)^{.}+\left(\frac{3\dot{B}}{AB}\right)^{2}
\right)\nonumber\\
&\times&\left(\frac{4B'}{{B^3r}}+2\frac{B''}{B^3}-\frac{B'^2}{{B^4}}
\right)+\frac{16}{3}\left(\left(\frac{3\dot{B}}{AB}\right)^{.}
+\frac{1}{2}\left(\frac{3\dot{B}}{AB}\right)^2\right)\nonumber\\
&\times&\left(-\frac{1}{B^2}\left(\frac{A'B'}{AB}\right)-\left(\frac{A'}{A}\right)^{'}
+\frac{2A'(Br)'}{AB^3r}\right).
\end{eqnarray}
It follows from Eq.(\ref{17}) that
\begin{equation}\label{23}
8{\pi}(q+q^G)B^2=\frac{2\Theta '}{3}.
\end{equation}
The Misner-Sharp (1964) mass becomes
\begin{equation}\label{24}
m(r,t)=\frac{r^3}{2}\left(\frac{B\dot{B}^2}{A^2}-\frac{B'^2}{B}-\frac{2B'}{r}\right)+\frac{s^2}{2rB}.
\end{equation}
We assume that the exterior region of the charged radiating star is
described by the charged Vaidya spacetime in a single null
coordinate
\begin{equation}\label{25}
{ds}^2=\left(1-\frac{M(\nu)}{\hat{R}}+\frac{Q^2}{\hat{R^2}}\right)d\nu^2+2d\nu
d\hat{R}-\hat{R}^2(d\theta^2+\sin^2 \theta d \phi^2),
\end{equation}
where $M(\nu),~Q^2$ and $\nu$ represent total mass, charge and
retarded time, respectively. For the matching of the interior and
exterior regions, we follow the procedure of Sharif and his
collaborators \cite{19}-\cite{25} and get
\begin{equation}
m(r,t)\overset{\Sigma}{=}M(\nu),\quad
p+p^G\overset{\Sigma}{=}(q+q^G)B ~\Leftrightarrow
s\overset{\Sigma}{=}Q.
\end{equation}
These are the necessary and sufficient conditions for the matching
of two regions of a collapsing star.

\section{Dynamical and Transport Equations}

In this section, we establish the dynamical and heat transport
equations for the collapsing charged radiating fluid. For this
purpose, we use the Misner and Sharp formalism \cite{9}. In this
case, the velocity of the collapsing matter is given by
\begin{equation}\label{26}
U=rD_tB,
\end{equation}
where $U<0$ for the collapsing fluid and
$D_t=\frac{1}{A}\frac{\partial}{\partial t}$. The Misner-Sharp mass
given by (\ref{24}) can be written as
\begin{equation}\label{27}
\frac{(Br)'}{B}=\left(1+U^2-\frac{2m(r,t)}{rB}+\left(\frac{s}{rB}\right)^2\right)^{\frac{1}{2}}=E.
\end{equation}
This is known as energy of the collapsing fluid. The proper time
derivative of the mass function (\ref{24}) leads to
\begin{eqnarray}\label{28a}
D_tm&=&r^3\frac{B\dot{B}\ddot{B}}{A^3}+\frac{r^3}{2}\left(\frac{\dot{B}}{A}\right)^3
-\frac{r^3B{\dot{B}}^2\dot{A}}{A^4}+\frac{r^3}{2}\frac{\dot{B}B'^2}{AB^2}
-r^3\frac{B'\dot{B'}}{BA}-\frac{r^2\dot{B}'}{A}\nonumber\\
&-&\frac{s^2\dot{B}}{2rAB^2}.
\end{eqnarray}
Using Eqs.(\ref{17}), (\ref{18}), (\ref{25}) and (\ref{26}), this
can be written as
\begin{equation}\label{28}
D_t m=-4\pi \left[{(p+p^G)U}+(q+q^G)BE
\right](rB)^2-\frac{s^2\dot{B}}{2rAB^2}.
\end{equation}
This equation gives the variation of the energy inside a gravitating
sphere of radius $R=rB$. As $U<0$, so the first term on right side
of the above equation increases the energy of a system while the
negative sign with second term implies out flow of energy.

In Misner-Sharp approach, the proper radial derivative is defined by
$D_R=\frac{1}{R'}\frac{\partial}{\partial r}$. The proper radial
derivative of Eq.(\ref{24}) with Eqs.(\ref{16}), (\ref{17}),
(\ref{25}) and (\ref{26}) leads to
\begin{equation}\label{31}
D_R
m=4\pi\left((\rho+\rho^G)+(q+q^G)B\frac{U}{E}\right)R^2+\frac{s}{R}D_R{s}-\frac{s^2}{2R^2}.
\end{equation}
The variation in energy between two adjacent layers of the fluid
inside the spherical boundary can be described by the above
equation. The contribution of electromagnetic field increases the
variation of energy inside the collapsing sphere. In this case, the
Gauss-Bonnet term affects the matter density and outward directed
heat flux. Since $q$ is directed outward due to $U<0$, so the
Gauss-Bonnet term causes the system to radiate away effectively.

The gravitational acceleration of the collapsing spherical boundary
can be obtained by using Eqs.(\ref{18}), (\ref{24}) and (\ref{26}),
which is given by
\begin{equation}\label{32}
D_tU=-\left(\frac{m}{(rB)^2}+4\pi(p+p^G)(rB)\right)+\frac{A'E}{AB}+\frac{s^2}{(rB)^3}.
\end{equation}
The simplification of
$\left(T^{\alpha\beta}+{T^{G}}^{\alpha\beta}\right)_{;\beta}\chi_{a}=0$,
for the given shearfree spherical boundary gives
\begin{eqnarray}\label{33}
&&\frac{1}{B} (p+p^G)'+\frac{A'}{BA}(\rho+\rho^G+p+p^G)
+(\dot{q}+\dot{q}^G)\frac{B}{A}+5(q+q^G)\frac{\dot{B}}{A}\nonumber\\&-&\frac{s\acute{s}}{4\pi
B R^4}=0.
\end{eqnarray}
Using the value of $\frac{A'}{A}$ from this equation in
Eq.(\ref{32}) with (\ref{24}) and (\ref{26}), we get the dynamical
equation
\begin{eqnarray}\label{34}
&&(\rho+\rho^G+p+p^G)D_tU=-(\rho+\rho^G+p+p^G)\left[m+4\pi(p+p^G)R^3-\frac{s^2}{R}\right]\frac{1}{R^2}\nonumber\\
&-&E^2\left[D_R(p+p^G)-\frac{s}{4\pi R^4}D_R s
\right]-E[5B(q+q^G)\frac{U}{R}+BD_t(q+q^G)].
\end{eqnarray}
This equation has form $force=mass~density \times acceleration$,
which is called "Newtonian" form of dynamical system, where mass
density $=\rho+\rho^G+p+p^G$.

According to this equation, the Gauss-Bonnet term affects the mass
density due to higher curvature terms. In this equation, the first
square bracket on right side is the gravitational force whose
Newtonian contribution is $m$ and relativistic contribution is
$p+p^G$. Thus the Gauss-Bonnet term affects gravitational force of
the gravitating sphere. The second term is the hydrodynamical force
which is affected by the electromagnetic field, overall this term
prevents collapse because $D_R(p+p^G)<0$ and $\frac{s}{4\pi R^4}D_R
s>0$. The last square bracket describes the role of heat flux during
the dynamics of collapsing sphere. Also, in this case, the first
term is positive ($U<0,~(q+q^G)>0$), indicating that outflow of heat
flux reduces the rate of collapse by producing radiations zone in
the exterior region of the collapsing sphere. The effects of
$D_t(q+q^G)$ can be explained by introducing the heat transport
equation as follows.

Here, we discuss the transportation of heat during the charged
shearfree collapse of radiative fluid in modified Gauss-Bonnet
gravity. To this end, we use Muller-Israel-Stewart heat transport
equation for heat conducting fluids \cite{28,29}. It is well-known
that Maxwell-Fourier law \cite{30} for the heat flux results to heat
equation which indicates perturbation at very high speed. For the
relativistic heat conducting charged fluids, the heat transportation
can be explained by Eckart-Landau theory \cite{31,32}, but results
obtained on the basis of this theory leads to some inconsistent
consequences. To resolve this problem, many relativistic theories
have been proposed. The common point of all these theories is that
these provide heat transport equation which is a hyperbolic
equation.

The heat transport equation in this case reads \cite{33}
\begin{eqnarray}\label{1a}
\tau h^{\mu \nu}V^\lambda
\tilde{q}_{\nu;\lambda}+\tilde{q}^\mu=\kappa h^{\mu
\nu}(T_{,\nu}-TA_\nu)-\frac{1}{2}\kappa T^2\left(\frac{\tau
V^\nu}{\kappa T^2}\right)_{;\nu}\tilde{q}^\mu,
\end{eqnarray}
where $h^{\mu\nu}$ is the projection tensor, $\tilde{q}=q+q^G$,
$\tau$ is relaxation time, $T$ is temperature and $\kappa$ is
thermal conductivity. For the under consideration spacetime, this
equation reduces to
\begin{eqnarray}\label{2a}
B\tau \frac{\partial}{\partial
t}{[({q}+{q}^G)B]}+(q+q^G)AB^2&=&-\kappa(TA)'-\frac{\kappa
T^2(q+q^G)B^2}{2}(\frac{\tau}{\kappa T^2}\dot{)}\nonumber\\
&-&3\frac{\tau \dot{B}B(q+q^G)}{2}.
\end{eqnarray}
Using Eqs.(\ref{26}), (\ref{27}) and (\ref{32}), this implies that
\begin{eqnarray}\label{3a}
&&BD_t(q+q^G)=-\kappa T \frac{D_t U}{\tau E}-\frac{\kappa
\acute{T}}{\tau B}-\frac{B(q+q^G)}{\tau}\left(1+\frac{\tau
U}{rB}\right)\nonumber\\
&-&\frac{\kappa T}{\tau
E}\left[m+4\pi{(p+p^G)R^3}-\frac{s^2}{R}\right]R^{-2}-\frac{\kappa
T^2 (q+q^G)B}{2\tau A}
\frac{\partial}{\partial t}\left(\frac{\tau}{\kappa T^2}\right)\nonumber\\
&-&\frac{3UB(q+q^G)}{2R}.
\end{eqnarray}
In order to study the effects of heat flux on the dynamics of
collapsing sphere in the modified Gauss-Bonnet gravity, we couple
the dynamical and heat transport equations given by Eqs.(\ref{34})
and (\ref{3a}). This coupling leads to
\begin{eqnarray}\label{3b}
&&(\rho+\rho^G+p+p^G)(1-\alpha)D_tU=F_{grav}(1-\alpha)+F_{hyd}+
\frac{E\kappa T'}{\tau B}\nonumber\\
&+&\frac{E(q+q^G)B}{\tau}+\frac{\kappa T^2(q+q^G)B}{2\tau A}
\frac{\partial}{\partial t}\left(\frac{\tau}{\kappa T^2}\right)
-\frac{5UB(q+q^G)E}{2R}.
\end{eqnarray}
Here
\begin{eqnarray*}
F_{grav}&=&-(\rho+\rho^G+p+p^G){\left(m+4\pi(p+p^G)R^3-\frac{s^2}{R}\right)}\frac{1}{R^2},\\
F_{hyd}&=&-E^2\left[D_R(p+p^G)-\frac{s}{4\pi R^4} D_R s\right],\\
\alpha&=&\frac{\kappa T}{\tau}(\rho+\rho^G+p+p^G)^{-1}.
\end{eqnarray*}

Now we explain the effects of the Gauss-Bonnet term and
electromagnetic field on the dynamics of the radiating collapsing
fluid. Since the Gauss-Bonnet term in the thermal coefficient would
affect the value of $\alpha$ and the factor $(1-\alpha)$ being the
the multiple of $F_{grav}$ would affect the value of $F_{grav}$. For
$\rho^G>0(<0)$ and $p^G>0(<0)$, the value of $\alpha$ in this case
will be less (more) as compared to GR and the value of $(1-\alpha)$
will be larger (smaller), consequently gravitational force will be
stronger. This fact can be explained by choosing a particular form
of $f(G)$ model, if $f(G)=f_0$, a positive (negative) constant,
Eqs.(\ref{20})-(\ref{22}) yield $\rho^G=-p^G= \frac{f_0}{2},~q^G=0$,
giving $\rho^G=\frac{f_0}{2}>0(<0)$ and the rate of collapse will be
increased (decreased). In both cases, the electromagnetic field
affects the gravitational and hydrodynamical forces of the system.

This fact can be verified for the more generalized $f(G)$ models.
When $\alpha\rightarrow1$, the system will be in hydrostatic
equilibrium as inertial and gravitational forces become zero in this
limit of $\alpha$. The amount of temperature $T$ for which
$\alpha\rightarrow1$ is equivalent to the expected amount of
temperature that can be achieved during the supernova explosion.
When $\alpha$ crosses the critical value, the gravitational force
plays the role of antigravity and the reversal of collapse would
occur.

\section{Summary}

In this paper, we have studied the dynamics of charged radiating
collapse in modified Gauss-Bonnet gravity. For this purpose, an
arbitrary function of Gauss-Bonnet invariant is introduced in the
Einstein-Hilbert. Generally, it is very difficult to deal with the
field equations in modified Gauss-Bonnet gravity. We use a
simplified form of the field equations formulated by Li et al.
\cite{26} using the CGI perturbation formalism with $3+1$ formalism.
In this approach, the nonlinear terms in $R_{\mu\nu}$ and
$R_{\mu\nu\lambda\sigma}$ can be expressed in the form of dynamical
quantities. The simplified field equations are used to study the
evolution of charged radiating star. The Darmois junction conditions
have been used to discuss the matching of the interior region of
collapsing star to the exterior charged Vaidya geometry. This
matching implies that on the boundary of collapsing star, the
effective pressure is balanced by the effective radial heat flux
provided the amount of charges in both regions of star is same. We
have formulated the dynamical equation which shows that Gauss-Bonnet
term affects the rate of collapse.

To discuss the transportation of heat during the charged radiative
fluid collapse, we have developed the heat transport equation for
the system under consideration. Further, to study the effects of
radial heat flux on the dynamics of charged collapsing fluid sphere,
we have coupled the dynamical equation with the heat transport
equation. The coupling of these equations leads to a single equation
({\ref{3b}}), which describes complete dynamics of the system. We
have investigated that when $\alpha$ exceeds the critical value, the
gravitational force plays the role of antigravity and the bouncing
would occur in the system.

The importance of charge term in the dynamics of dissipative
collapse has been discussed. In particular, it is worth mentioning
that unlike pressure "charge term" does not act as regeneration
source. In other words, electric charge does not increase the active
gravitational mass. On the basis of this fact, we can conclude that
presence of charge is important for the dynamics of dissipative
collapse. A model with bouncing behavior has been presented
numerically by Herrera et al. \cite{35}.  This work can be extended
for the generalized $f(G)$ model \cite{8a} like $f(G)=AG+
BG^{\frac{1+\beta}{4{\beta}^2}}$, $A$, $B$ and $\beta$ are
constants.


\vspace{0.25cm}
\begin{thebibliography}{40}
\bibitem{1a} S. Nojiri and S.D. Odintsov: Phys. Rev.
 \textbf{D 75}, 086005(2006).

\bibitem{1b} P.T. Sotiriou and S. Liberati: Ann. Phys.
 \textbf{322}, 935(2007).

\bibitem{1c} P.T. Sotiriou and V. Faraoni: Rev. Mod. Phys. \textbf{82}, 451(2010).

\bibitem{1d} A. De Felice and S. Tsujikawa:
Living Rev. Rel. \textbf{13}, 3(2010).

\bibitem{2a} R.G. Bengochea and R. Ferraro: Phys. Rev. \textbf{D79}, 124019(2009).

\bibitem{2b} E. V. Linder: Phys. Rev.
\textbf{D81}, 127301(2010).

\bibitem{2c} A. De Felice and S. Tsujikawa: Phys.
Lett. \textbf{B675}, 1(2009).

\bibitem{3} T. Harko, F.S.N. Lobo, S. Nojiri
and S.D. Odintsov: Phys. Rev. \textbf{D84}, 024020(2011).

\bibitem{3a} E.J. Copeland, M. Sami and S. Tsujikawa: Int. J. Mod. Phys. \textbf{D15},
1753(2006).

\bibitem{3b} S. Capozziello: Int. J. Mod. Phys. \textbf{D11}, 483(2002).

\bibitem{4} S. Nojiri and S.D. Odintov:
 Phys. Lett. \textbf{B631}, 1(2005).

\bibitem{5} S. Nojiri, S.D. Odintov and O.G. Gorbunova:
 J. Phys. \textbf{A39}, 6627(2006).

\bibitem{6} G.Cognola, E. Elizalde, S. Nojiri, S.D. Odintov and
S. Zerbini: Phys. Rev. \textbf{D75}, 086002(2007).

\bibitem{7} M. Gasperini
and G. Veneziano: Astropart. Phys. \textbf{1}, 317(1993).

\bibitem{8}  G.Cognola, E. Elizalde, S. Nojiri, S.D. Odintov and
S. Zerbini: Phys. Rev. \textbf{D73}, 084007(2006).

\bibitem{8a} R. Myrzakulov, D. S$\acute{a}$ez-G$\acute{o}$emz and A. Tureanu:
Gen. Relativ. Gravit. \textbf{43}, 1671(2011).

\bibitem{9} C.W. Misner and D.H. Sharp: Phys. Rev.
\textbf{136}, B571(1964).

\bibitem{10} C.W. Misner: Phys. Rev. \textbf{137}, B1360(1965).

\bibitem{11} D. Kazanas and D. Schramm: \textit{Sources of Gravitational
Radiation} (Cambridge University Press, 1979).

\bibitem{12} L. Herrera, A. Di
Prisco, J. Martin, J. Ospino, N.O. Santos and O. Troconis: Phys.
Rev. \textbf{D69}, 084026(2004).

\bibitem{13} A. Mitra: Phys. Rev. \textbf{D74}, 024010(2006).

\bibitem{14} L. Herrera and N.O. Santos: Phys. Rev.
\textbf{D 70}, 084004(2004).

\bibitem{15} R. Chan: Astron. Astrophys. \textbf{368}, 325(2001).

\bibitem{16} L. Herrera: Int. J. Mod. Phys. \textbf{D15}, 2197(2006).

\bibitem{17} L. Herrera, A. Di Prisco, E. Fuenmayor and O. Troconis: Int.
J. Mod. Phys. \textbf{D18}, 129(2009).

\bibitem{18} A. Di Prisco, L. Herrera, G. Denmat,  M.A.H. MacCallum and
N.O. Santos: Phys. Rev. \textbf{D76}, 064017(2007).

\bibitem{19} M. Sharif and G. Abbas: Astrophys. Space Sci. \textbf{335}, 515(2011) .

\bibitem{19a} M. Sharif and G. Abbas: Astrophys. Space Sci. \textbf{327}, 285(2010).

\bibitem{23} M. Sharif and G. Abbas: Mod. Phys. Lett. \textbf{A4},
2551(2009).
\bibitem{23a} M. Sharif and G. Abbas: J. Korean Phys. Soc. \textbf{56},
529(2010).
\bibitem{23b} M. Sharif and G. Abbas: Gen. Relativ. Gravit. \textbf{44},
2353(2010).
\bibitem{24} M. Sharif and G. Abbas: J. Phys. Soc. Jpn. \textbf{80}, 104002(2011)104002.
\bibitem{25} M. Sharif and G. Abbas: J. Phys. Soc. Jpn. \textbf{81}, 044002(2012).
\bibitem{25} M. Sharif and G. Abbas: J. Phys. Soc. Jpn. \textbf{82}, 034006(2013).

\bibitem{26} B. Li, J.D. Barrow and F.D. Mota: Phys. Rev. \textbf{D76}, 044027(2007).

\bibitem{27} G. Darmois:
\textit{Memorial des Sciences Mathematiques }(Gautheir-Villars,
Paris, 1927).

\bibitem{28} I. M$\ddot{u}$ller: Z. Physik \textbf{198}, 329(1967).

\bibitem{29} W. Israel and J. Stewart: Phys. Lett. \textbf{A 58}, 2131(1976).

\bibitem{30} D. Pav\'{e}n, D. Juo and M. Casas-V\'{a}zques: Ann. Inst.
Henri Poincar\'{e} \textbf{A36}, 79(1982).

\bibitem{31} C. Eckart: Phys. Rev. \textbf{58}, 919(1940).

\bibitem{32} L. Landau and E. Lifshitz: \textit{Fluid Mechanics} (Pergamon
Press, 1959).

\bibitem{33} J. Martinez: Phys. Rev. \textbf{D53}, 6921(1996).

\bibitem{35} L. Herrera, A. Di Prisco and W. Barreto: Phys. Rev. \textbf{D73}, 024008(2006).

 \end{thebibliography}
\end{document}